# Co-word Maps and Topic Modeling:

# A Comparison Using Small and Medium-Sized Corpora (*n* < 1000)

Loet Leydesdorff *[a] & Adina Nerghes [b]

**Abstract**

Induced by "big data," "topic modeling" has become an attractive alternative to mapping co-words in terms of co-occurrences and co-absences using network techniques. Does topic modeling provide an alternative for co-word mapping in research practices using moderately sized document collections? We return to the word/document matrix using first a single text with a strong argument ("The Leiden Manifesto") and then upscale to a sample of moderate size (*n* = 687) to study the pros and cons of the two approaches in terms of the resulting possibilities for making semantic maps that can serve an argument. The results from co-word mapping (using two different routines) versus topic modeling are significantly uncorrelated. Whereas components in the co-word maps can easily be designated, the topic models provide sets of words that are very differently organized. In these samples, the topic models seem to reveal similarities other than semantic ones (e.g., linguistic ones). In other words, topic modeling does not replace co-word mapping in small and medium-sized sets; but the paper leaves open the possibility that topic modeling would work well for the semantic mapping of large sets.

[a] University of Amsterdam, Amsterdam School of Communication Research (ASCoR), PO Box 15793, 1001 NG Amsterdam, The Netherlands; loet@leydesdorff.net ; * corresponding author.
[b] VU Amsterdam, Department of Organization Sciences, De Boelelaan 1081, Amsterdam 1081 HV, The Netherlands; adina.nerghes@vu.nl

## 1. Introduction

In recent years, the number of options for generating semantic or co-word maps from texts has proliferated. Referred to as maps (Carley, 1997b), semantic networks (Lehmann, 1992), networks of concepts (Popping, 2000), or networks of words (Danowski, 1993), this approach to the analysis of texts assumes that the content of communication in language can be modeled as networks of words and the relations among them (Danowski, 2009; Sowa, 1992). More fundamentally, Hesse (1980)—following Quine's (e.g., 1960) philosophy of science—argued that networks of co-occurrences and co-absences of words are shaped at the epistemic level and can thus reveal the evolution of the sciences in considerable detail (Kuhn, 1984; cf. Kuhn, 2000, p. 2). From this perspective, the latent structures in the networks can be considered as the organizing principles or the codes of the communication (Luhmann, 1990; Rasch, 2002). In principle, this "linguistic turn in the philosophy of science" (Rorty, 1992 [1967]) makes the sciences amenable to measurement and sociological analysis (Law & Lodge, 1984; Leydesdorff, 2007).

In science and technology studies (STS), Callon *et al*. (1983) was the first to place the development of co-word maps on the research agenda, but the development of software for the mapping remained slow during the 1980s (Leydesdorff, 1989; Tijssen & van Raan, 1989). At the time, much of the visualization still had to be done by hand (e.g., Rip & Courtial, 1984; cf. Leyesdorff, 1992). Co-word mapping using programs freely available on the internet began to take shape only in the second half of the 1990s, when network visualization programs adapted to the graphical user interfaces of new versions of Microsoft's Windows and the Apple Macintosh became available.

One of us, for example, developed a series of routines during the 1990s that allowed, among other things, for semantic mapping using text files as input and Pajek for the visualization.[1] At that time, two programs—ti.exe at http://www.leydesdorff.net/software/ti and fulltext.exe at http://www.leydesdorff.net/software/fulltext—were hugely in demand because of a lack of alternatives (Vlieger & Leydesdorff, 2011; Leydesdorff & Welbers, 2011). Nowadays, however, one has a plethora of options, and the user may find it difficult to choose which representation to use. Among others, let us mention: Automap in conjunction with ORA (Carley *et al*., 2013, Carley *et al*., 2013b),[2] VOSviewer (van Eck *et al*., 2010),[3] ConText (Diesner *et al*., 2015),[4] and Wordjj.exe (Danowski, 2009).[5] Whereas the purpose of the different programs is the same—visualization of the latent structures in the data (Lazarsfeld & Henry, 1968)—their results can be very different. Two parameter choices are relevant: similarity criteria and clustering algorithms (e.g., Leydesdorff, 2008; Van Eck & Waltman, 2009).

Our objective in this study is not to compare these word-mapping programs with one another. Rather, we aim to compare (co-)word maps with the results of a more recent approach called "topic modeling" (Blei *et al*., 2003). Whereas co-word mapping originated in STS and was further developed in library and information science, topic modeling found its origins in natural language processing and the computer sciences. Using a probabilistic algorithm, the program searches for a specified number of topics indicated by sets of words extracted from the texts under study. The output allows for inspection of both documents in terms of their participation in the various topics and the relative weights of each word in the topics. The topic model provides an orientation for organizing large collections of documents. Mapping and overlays to maps are not a primary objective of topic modeling; but this extension has recently been explored using VxOrd as an algorithm for mapping large data sets (Klavans & Boyack, 2014; Talley *et al*., 2011).

---

[1] Pajek is a network analysis and visualization program freely available for non-commercial usage at http://mrvar.fdv.uni-lj.si/pajek/ (Retrieved on October 13, 2015).
[2] Available at http://www.casos.cs.cmu.edu/projects/automap/ and http://www.casos.cs.cmu.edu/projects/ora/ (Retrieved on October 13, 2015).
[3] Available at http://www.vosviewer.com (Retrieved on October 13, 2015).
[4] ConText provides also the option of topic modeling. ConText is available at http://context.lis.illinois.edu/ (Retrieved on October 13, 2015).
[5] Available at http://wordij.net/ (Retrieved on October 13, 2015).

In a study for the U.S. National Institute of Health (NIH), for example, Talley *et al*. (2011) developed a topic model for the accumulated grants (70,000 – 80,000 per year) and the ~220k MEDLINE journal articles published between 2007 and 2009, that cited NIH grants "to enhance the statistical robustness of the analysis for the corresponding areas of research" (Supplementary Methods, p. 18). The testing and validation of a topic model are no *sinecures*: a topic model needs to be trained on a subset and a number of parameters have to be chosen such as the number of topics to be distinguished (T), the concentration of topics in a document (alpha), and terms in a topic (beta). The results have to be screened manually and/or by automatic scoring systems because topics of poor quality remain possible, also after a number of iterations.

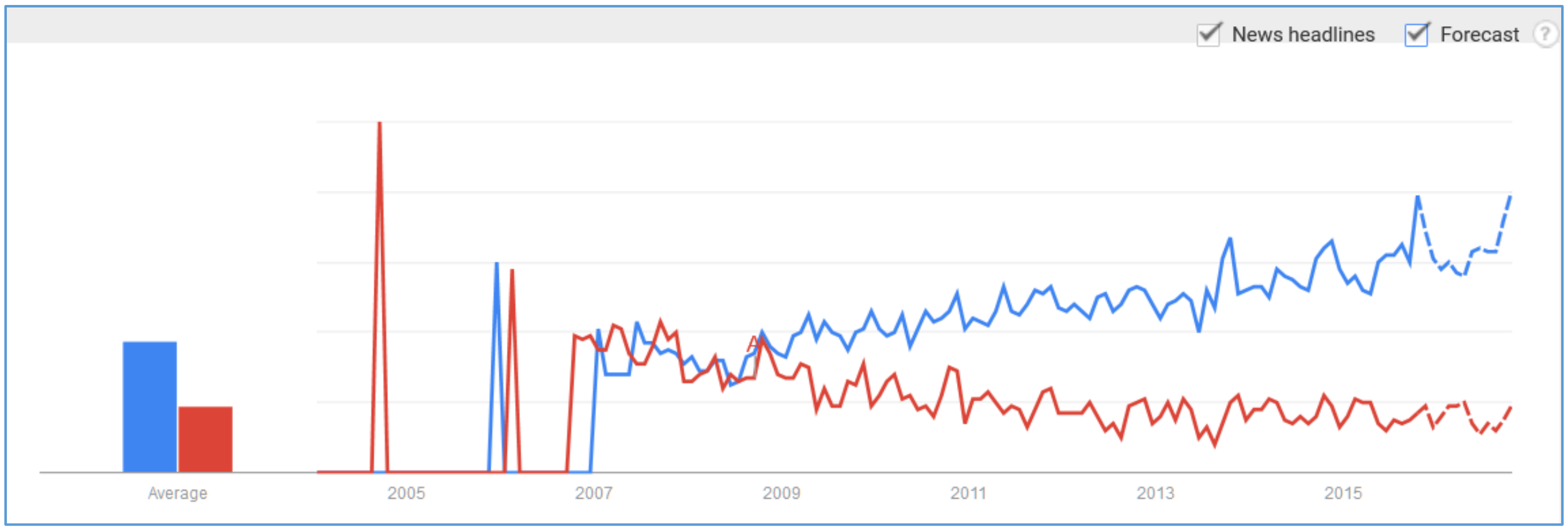

**Figure 1**: Google Trends for "topic model" (blue) and "semantic map" (brown) on November 1, 2015.

Topic modeling has become increasingly popular in recent years. Figure 1 shows the results of comparing the phrase "topic model" (in blue) with "semantic map" (brown) using Google Trends (on November 1, 2015). The figure indicates that topic models are currently more popular than semantic maps for the purpose of summarizing corpora of texts. This has been enhanced by the availability of a user-friendly routine for topic modeling at [https://code.google.com/p/topic-modeling-tool/](https://code.google.com/p/topic-modeling-tool/) since 2013. This routine provides a graphical user interface to MALLET (the common name of the "MAchine Learning for LanguagE Toolkit") which has been developed by a team at UMass Amherst since 2002 (McCallum, 2002).[6]

MALLET has set the standard for topic modelling ever since its introduction (Chang *et al.*, 2009). Among the many algorithms developed for topic modeling, the original model (Latent Dirichlet Allocation or LDA) has remained the most frequently used and was also selected for the new application. The same application—that is, LDA in MALLET—is also available in other programs such as Jana Diesner's program ConText (Diesner *et al.*, 2015). As Jacobi *et al.* (2015) formulate: "LDA is a cutting edge technique for content analysis, designed to automatically organize large archives of documents based on latent topics—measured as patterns of word (co-)occurrences." The National Science Foundation of the U.S.A., for example, used a similar topic model based on LDA for the internal categorization of its awards (Nichols, 2014).

[6] MALLET is available at [http://mallet.cs.umass.edu](http://mallet.cs.umass.edu) . The topic-modelling tool at [https://code.google.com/p/topic-modeling-tool/](https://code.google.com/p/topic-modeling-tool/) provides an interface to MALLET with the same functionality.

The problem with analyzing large corpora of texts that are beyond the human capacity to comprehend (by reading) remains the validity of the results. Words are so flexible that one can almost always provide an interpretation of groups of words *ex post*. In this study, we reverse the reasoning by taking a bottom-up perspective: we compare the two methods step-by-step by using first a single document and then a set of 687 documents. We begin by making the word/document matrix using our own routines so that we are able to control the steps. Our methods are interfaced with Pajek and SPSS for the statistical analysis. Visualizations in Pajek will be exported to VOSviewer for presentational purposes. From the perspective of more traditional statistics, one can compare the matrices (e.g., excel sheets) resulting from the topic modeling with the factor loadings and factor scores of the word/document matrix. However, the algorithm is very different: topic modeling can be considered as belonging to another generation of statistics since one can organize huge amounts of data (in this case, texts). Ding (2011) compared the topical model with the topological one in community-detection algorithms (Blondel et al., 2008; Newman & Girvan, 2004), and specified the differences.

For the purpose of visually inspecting the differences and similarities of the two representations, the results of the topic modeling can be overlaid on the semantic maps. However, for more robust and precise tests of similarity between the results of the two methods, one may use chi-square (in SPSS).[7] In this paper, we use Cramér's *V* because it provides a value of the association between zero and one, and also tests for significance. We background the technical details (e.g., Chang *et al*., 2009; Grimmer & Stewart, 2013; Mohr & Bogdanov, 2013), and focus pragmatically on comparing the two approaches from a user's perspective, with a specific interest in small and medium size data sets.

We are aware that our design begins with the co-word model and, hence, the co-word map becomes the baseline for the assessment of the topics. The maps, however, are illustrations meant to show the effects of the differences and not meant as Procrustes beds. For this purpose, we added statistics. Additionally, the comparison is partly hindered by differences in input and output required by or resulting from the two models. For example, the clustering algorithm in Pajek we use (Blondel *et al*., 2008), attributes each word to a single group, whereas each word is considered as a multinominal distribution over topics in the results of a topic model, and can thus be attributed to more than a single topic. In this case, we try to give the results of the topic model priority and compare with the single attribution of this word in the co-word model in each instance. However, we do not train the model or revisit it—as one should—but remain on the user side by taking up the invitation of the topic-modeling tool at the Google site. Furthermore, an *ex ante* choice has to be made for the number of topics; we orient ourselves using the co-word model and factor analysis of the word/document matrix. However, we also follow the suggestion of one of the referees to explore ten topics in addition to the five suggested by the multi-variate analysis of the matrix. The visuals are used didactically for illustrating the analytical reasoning.

[7] In a similar vein, Rafols & Leydesdorff (2009) compared content-based classifications of journals with the results of algorithmic decompositions in terms of communities (e.g., Newman & Girvan, 2004).

## 2. Data and methods

In order for us to interpret and perhaps validate the results, we first reduce the problem to the analysis of a single text with which users may be familiar (or can easily become so) because it is at the heart of our field. For this purpose, we chose the so-called "Leiden Manifesto" that was published by a group of colleagues in *Nature* on April 23, 2015 (Hicks *et al*., 2015). The Manifesto intends to provide guidelines for the use of metrics in research evaluation. As of this date, the Manifesto has been translated into nine languages other than English (at http://www.leidenmanifesto.org/translations.html ), and was cited 25 times in five months (at Google Scholar, 28 September 2015). We analyze the Leiden Manifesto in terms of the 26 substantive paragraphs (as units of analysis) in which the argument is made.

As noted, topic modeling is not developed for a single text. Generally, topic models are used for the analysis of large corpora of documents. As a step in this direction, we upscale as a second example to a moderately sized set of 687 documents retrieved using ‘"Leiden ranking" OR "Leiden rankings"‘ as a search string in Google Scholar (on October 4, 2015). The search string "Leiden Manifesto" provided us first with only 53 records at this date, of which many are substantially consensual, whereas the "Leiden Rankings" are more controversial. The first edition of the Leiden Rankings was from 2007 (Waltman *et al*., 2012, at p. 2420), and thus this discussion has had more time to accumulate.

The idea behind “topic modeling,” of course, is to analyze much larger sets that cannot possibly be read by human readers. However, the quality of the representation cannot easily be controlled when using such “distant reading” (Moretti, 2013). Beyond hundreds of documents or topics, our control as human readers diminishes rapidly. The larger set of 687 cases scales the smaller set of 26 paragraphs in a single document by more than an order of magnitude, and therefore seems a prudent choice given the objectives of this study. The number also conveniently remains below the display limit of Google at 1000 records.[8]

Given our user orientation, we select default values within the programs unless we have specifiable reasons for choosing otherwise. We discuss the specific choices in each program in relation to discussing the results. For example, determining the number of topics in a topic model is not a sinecure: Blei *et al.* (2003) developed “perplexity” as a measure for this purpose.[9] One can calculate the perplexity by training the model on a subset of the data; but this is beyond the practice of an average user who is prompted by the online program to specify the number of topics. We will conjecture this number based on the results of a community-finding algorithm (Blondel *et al*., 2008), of the co-word map, and the factor analysis of the word-document matrix. These choices make our results comparable across the two types of analysis, given that we also use the same samples as inputs. However, our results may be biased in favor of co-word modeling because we did not test the topic model independently. We return to this issue of our lens in the conclusions section.

[8] On the same date, the search string ‘"ARWU ranking" OR "ARWU rankings" OR "Shanghai ranking" OR "Shanghai rankings"‘ returned 2020 results.

[9] The lower the perplexity, the better the prediction. However, the best fit does not have to accord with the optimal interpretability of the topics (Chang *et al*., 2009; Jacobi *et al*., 2015).

## 3. Results

### *3.1. The Leiden Manifesto*

After removing para-textual information such as headings, references, etc., the Leiden Manifesto contains 26 substantive paragraphs that shape the argument using 724 unique words occurring 1,926 times. By applying a list of 429 stop words available at http://www.lextek.com/manuals/onix/stopwords1.html, 550 unique words remain, of which 75 occur more than twice. We used these 75 words for the analysis and generated the word/occurrence matrix with the words as column variables and the 26 paragraphs as rows.[10] In order to optimize the visualization, we normalized the word vectors using the cosine (Salton & McGill, 1983).[11] A threshold level is chosen pragmatically at cosine > 0.2; the size of the nodes is proportional to the logarithm of the word frequencies.

[10] We also equate plural and singular nouns. The analysis is based on ti.exe available at http://www.leydesdorff.net/software/ti .

[11] The choice of the cosine as a similarity criterion is motivated by our wish to remain as close as possible to the representation of factor-analytic results, which are based on Pearson correlations. The cosine can be considered as a non-parametric variant of the Pearson correlation since the normalization to the mean is omitted (Ahlgren *et al.*, 2003; Leydesdorff, 2008; cf. Van Eck & Waltman, 2009).

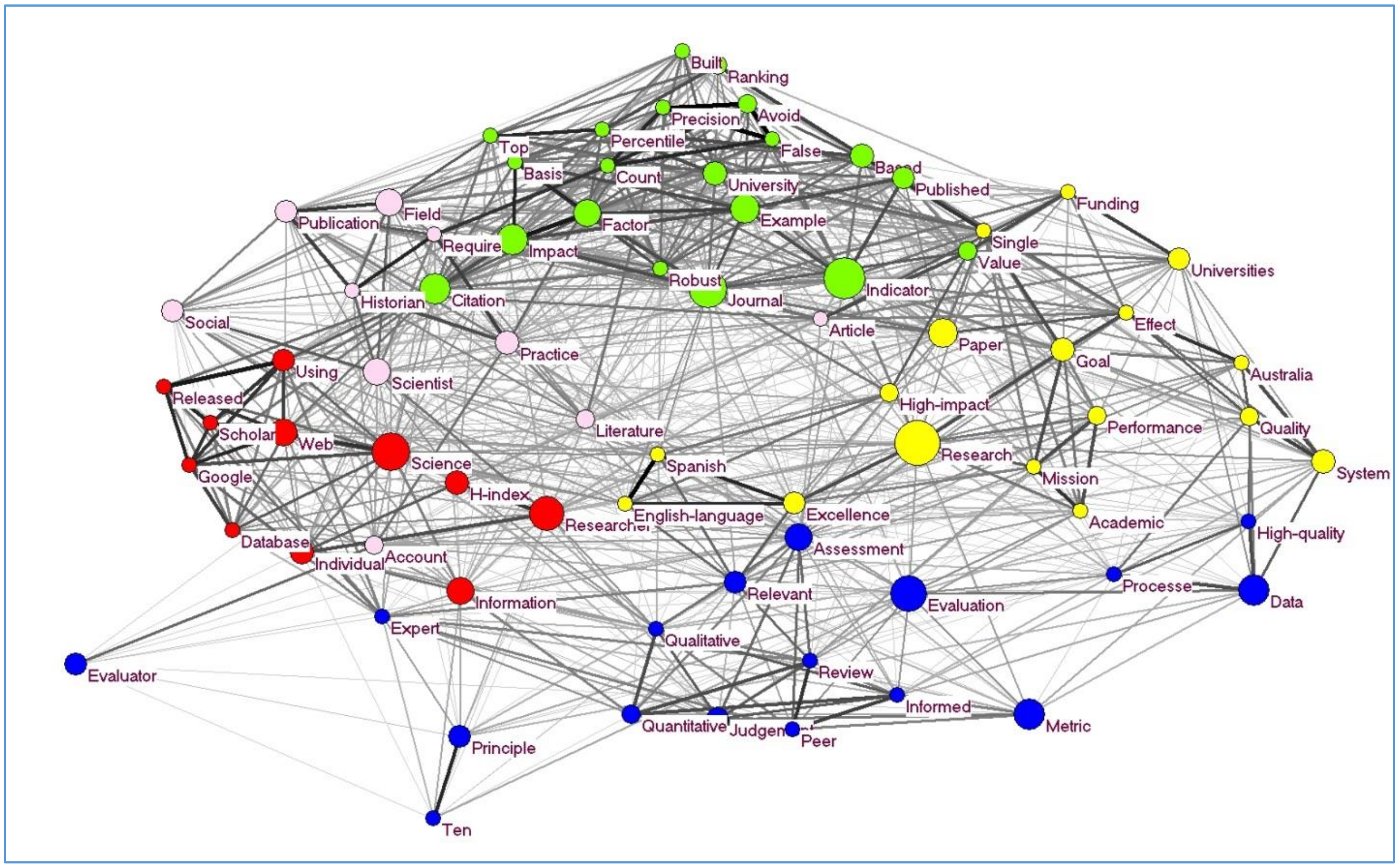

**Figure 2**: Five clusters of 75 words in a cosine-normalized map (cosine > 0.2) distinguished by the algorithm of Blondel *et al*. (2008); modularity *Q* = 0.27. Kamada & Kawai (1989) used for the layout.

Figure 2 shows a map of the 75 words using the layout of Kamada & Kawai (1989) as implemented in Pajek. Using the implementation of Blondel *et al*.'s (2008) algorithm in Pajek, five clusters are robustly distinguished (*Q* = 0.27). The green-colored cluster at the top of the figure, for example, refers to university ranking, whereas the blue one at the bottom refers to evaluation and peer review.

**Table 1**: Highest (and lowest) factor loadings in the case of a five-factor solution (Varimax rotated) of the word/document matrix.

| **Component 1** | **Component 2** | **Component 3** | **Component 4** | **Component 5** |
|---|---|---|---|---|
| false (.903)<br>avoid (.903)<br>precision (.903)<br>count (.868) | percentile (.887)<br>top (.761)<br>based (.735)<br>university (.698) | Google (.941)<br>released (.875)<br>scholar (.848)<br>web (.828) | publication (.645)<br>historian (.574)<br>article (.569)<br>field (.548) | effect (.773)<br>goal (.760)<br>Australia (.639)<br>quality (.620) |
| | | | *informed (-.624)*<br>*quantitative (-.614)*<br>*peer (-.579)*<br>*judgement (-.509)* | |

Forcing a five-factor solution (Varimax rotated) on the underlying word/document matrix provides us with guidance on how to designate the clusters; but a comparison of Table 1 with the groupings in Figure 2—based on a community-finding algorithm—teaches us that there remains room for divergent interpretations. The grouping using the cosine is also a bit different because the factor analysis uses the Pearson correlation (Ahlgren *et al*., 2003). In sum, both similarity measures and clustering algorithms affect the grouping.

The first five factors explain only 47.04% of the variance in the matrix. This relatively low percentage is typical for word/document matrices: there is structure in language, but it is not so pronounced as to be obvious in terms of word and co-word distributions (Leydesdorff, 1997). Words may have different meanings in different contexts, although one expects the meaning of specific words to be stable within a single text (Leydesdorff & Hellsten, 2006).

The algorithm of VOSviewer tends to generate more clusters than the Blondel-algorithm. Indeed, a sixth group is sometimes indicated with the words “Spanish,” “English-language,” “Excellence,” and “High-impact”; but in other runs one can also reproduce the option of precisely these *same* five clusters. The modularity parameter *VOS* is 0.38. Within VOSviewer, however, one has other options for the visualization, such as the cluster density view or hotmap. Among other things, VOSviewer has the advantage of optimizing the labels in the visualization so that they are not blurred. When using a semantic map in an argument, however, this is not always an advantage because one may wish to discuss a specific word in its context, which for one reason or another, may be difficult to foreground.

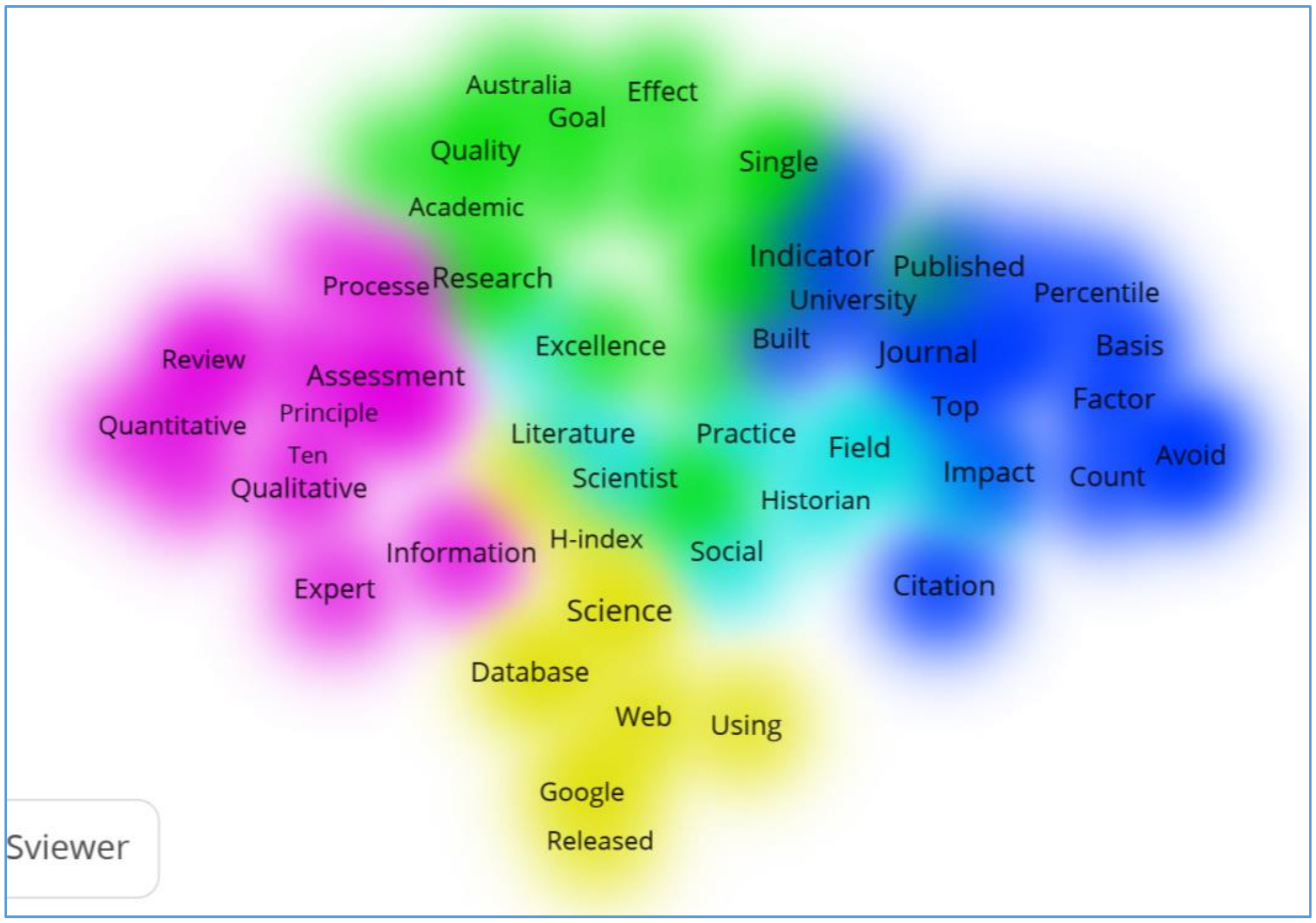

**Figure 3**: Same clusters as in Figure 2, but using the layout of VOSviewer.

Figure 3 shows the same data as in Figure 2, but using the layout of VOSviewer instead of Kamada & Kawai (1989). Pajek—and also Gephi[12]—allows one to choose among the different layout algorithms, and only thereafter decide to export to VOSviewer for a richer visualization. However, one can also export to other formats (such as .svg or .eps) and embellish the representation in greater detail and using one's own preferences (de Nooy *et al*., 2011; Vlieger & Leydesdorff, 2011). The graphical quality of the map can be important when the map is used as a persuasive illustration in an argument.

[12] Available at http://gephi.github.io/ (retrieved on October 28, 2015).

Alternatively, one can feed the textual data also directly into VOSviewer. VOSviewer prompts for a few choices. In this case, we chose for "full counting" since we have numerical (counting) values in the data. The program recognized 370 terms in the data, of which 31 occur three or more times. Thereafter, VOSviewer suggests (default) applying a relevance score for selecting the 60% most relevant terms, in this case 19 terms. Given our design, however, we preferred to include all 31 terms.

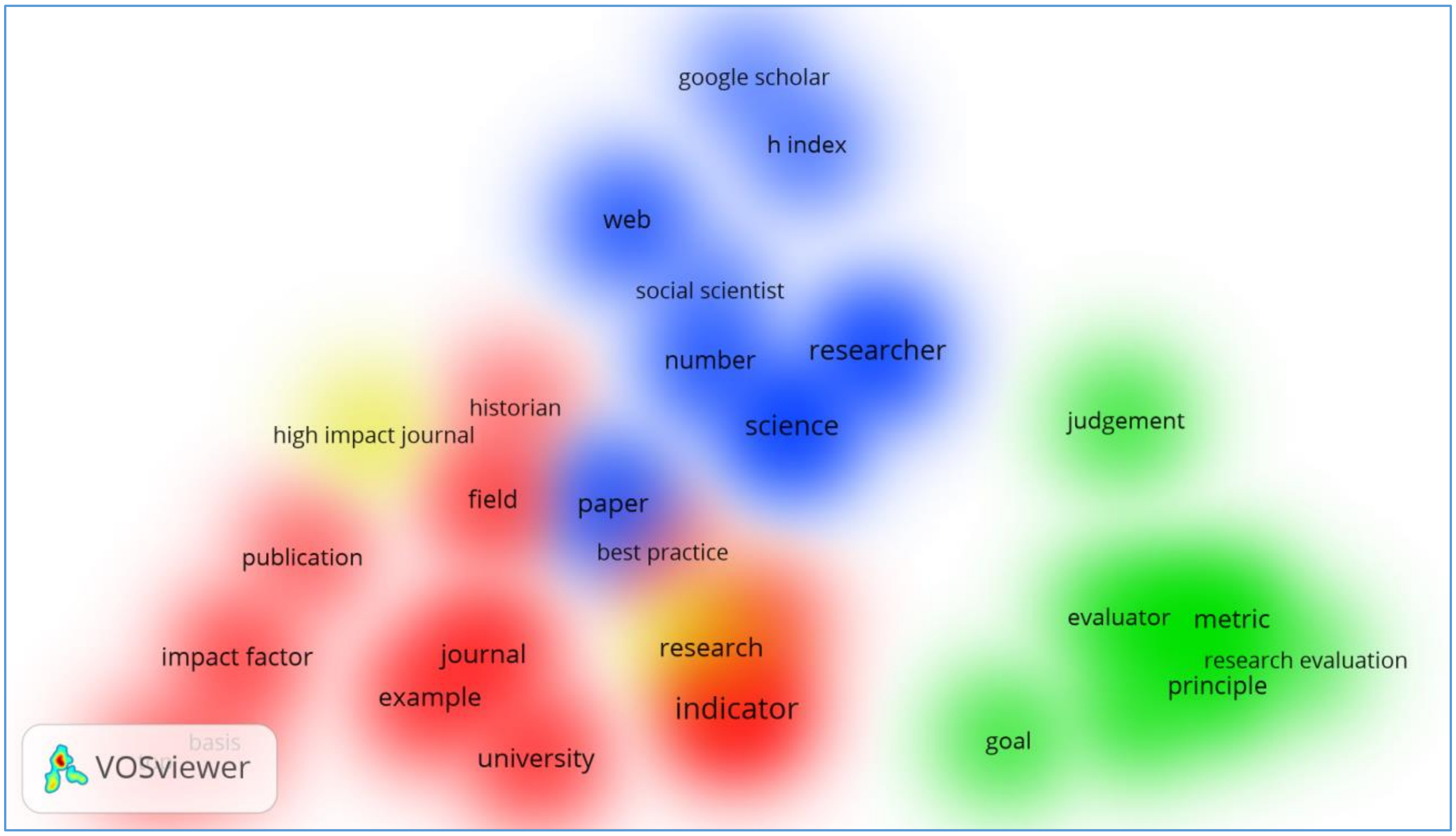

**Figure 4**: Thirty-one words organized into four clusters by VOSviewer.

The smaller set of words (31 as against 75) leads to a smaller number of clusters (four instead of five; Figure 4). The fourth cluster consists of only two (dispersed) words: "research" and "high-impact journal". Van Eck *et al.* (2010; van Eck & Waltman, 2011) explain the linguistic filter used in VOSviewer to compose noun phrases such as "high-impact journal" as terms instead of single words. The user has the option to select the words to be included in the analysis manually.

### *3.2 A topic model of the Leiden Manifesto*

The topic-modeling tool available at https://code.google.com/p/topic-modeling-tool/ uses Latent Dirichlet Allocation (LDA). The mathematics of LDA is complex (Blei *et al*., 2003). The user has to choose the number of topics to be extracted from the data. As a default, each topic is represented in the output by ten words. Let us for the sake of comparison extract five topics, include the same list of stop words, and otherwise use the default values for the (advanced) options. Table 2 provides the relevant results (word lists) of this model.

**Table 2**: Word lists of five topics in the Leiden Manifesto, using LDA/MALLET as implemented at https://code.google.com/p/topic-modeling-tool/

| *topicId* | *words..* |
|---|---|
| 1 | science evaluation web scientists index factor top peer journal judgement |
| 2 | research indicators based indicator assessment evaluators making effects built excellence |
| 3 | data quality published field metrics avoid precision false robust systems |
| 4 | journals high papers language number social english review national spanish |
| 5 | impact researchers citation information university relevant universities individual practice literature |

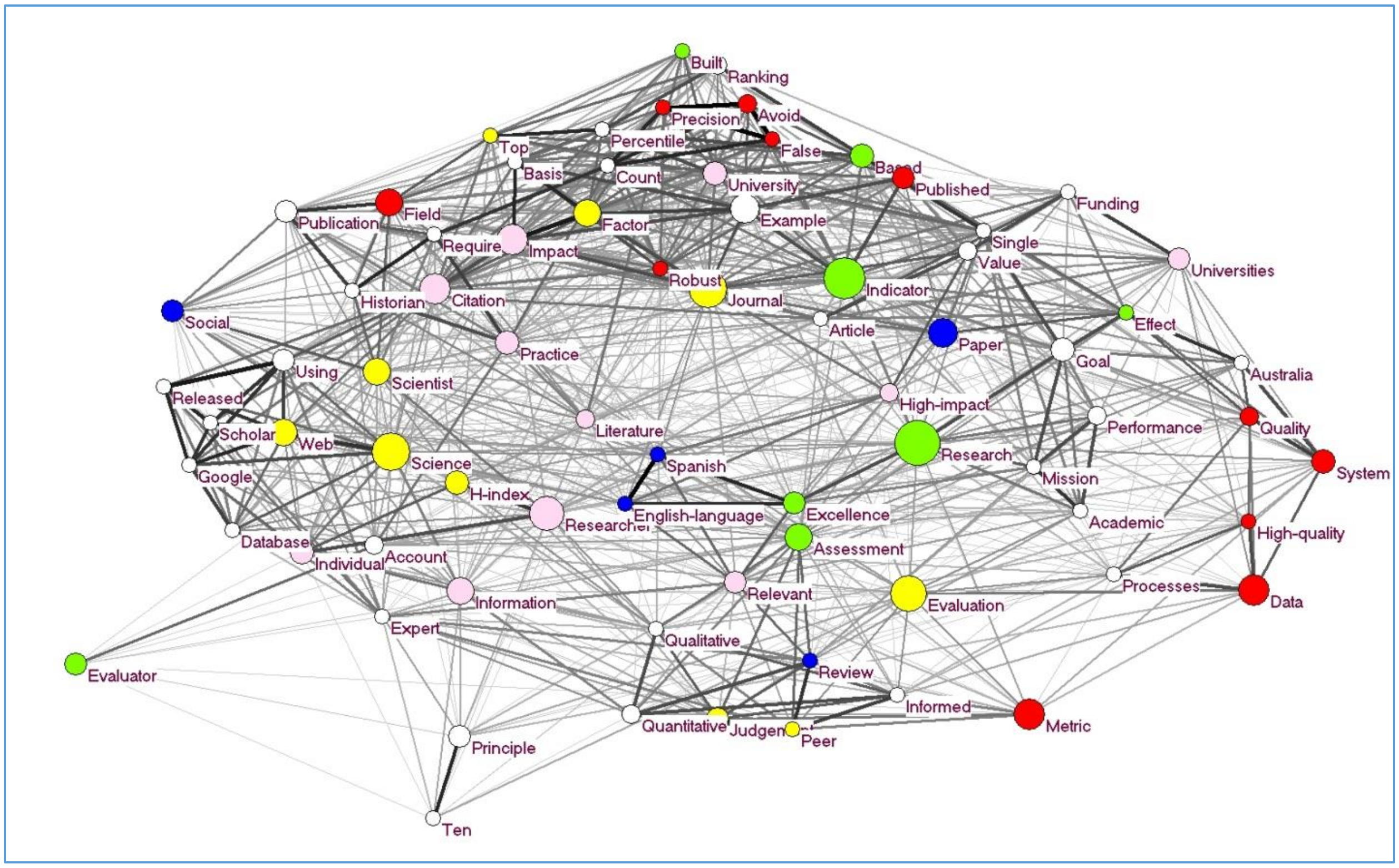

**Figure 5**: Layout as in Figure 2; but nodes are colored according to the LDA model. (Words not covered by the LDA output are colored white.)

Figure 5 shows a structure as in Figure 2, but this time the nodes are colored according to the topical assignment of each word (as shown in Table 2). Because we used 75 words above, and the five topics are limited to ten words each, 25 words (75 – (5 * 10)) remain unassigned. In some cases, there is no one-to-one match between our 75 words and the 50 words included in the topics because the topic modeler handles compound words like “high-impact” as two individual words: “high” and “impact”. (Here we used the word “impact” for the assignment in the map.) In sum, we were able to classify 45 of the 75 words and color them as shown in Figure 5. However, among the thirty words indicated not thus classified (white vertices) are words such as "principles", "performance", "mission", "academic", "publication", etc., all of which belong to the core of the argument in the Leiden Manifesto.[13]

The clustering in the 45 matched words has a Cramér's $V = .311$ ($p = .359$) with the clustering underlying Figure 2; and Cramér's $V = .313$ ($p = .344$) with the VOSviewer classification used for Figure 3, whereas the VOSviewer's and Blondel *et al.*'s (2008) algorithms lead to two highly correlated clusterings (Cramér's $V = .959$; $p < .000$). Thus, the topic model is significantly different in all respects from the maps based on co-occurrences of words. Note that this does not imply that the topic model is "wrong," but only that the results are incompatible with those of the co-word map, in the case of a corpus of 26 paragraphs constituting a single text. One cannot exclude that the grouping of the documents over the topics is meaningful despite the relative incomprehensibility of the assignment of words to topics. We tested this by comparing the 26 paragraphs in terms of the sequence number of the topic with the highest contribution, on the one side, and the sequence number of the factor with the highest absolute factor score. This relation is significant at the 5% level (Cramér's $V = .517$; $p = 0.034$).

In summary: the main difference between using VOSviewer or our routines with Pajek for the mapping is that VOSviewer pre-processes the data in search of noun-phrases. The other selections made by VOSviewer are suggestions, which can be declined. Furthermore, one is bound to the layout of VOSviewer, but the files can be prepared and saved in Pajek format and then represented with other layouts and analyzed in statistical terms. The philosophy, however, is different: VOSviewer is organized from the perspective of visualization, whereas in Pajek the emphasis is on the network statistics; visualization follows as a second objective. While the results using these two methods for making co-word maps are recognizably similar and significantly correlated statistically, the results of the topic model were significantly non-correlated and not easy to interpret.

[13] Using MALLET, all words are multinominally distributed over the topics, but this matrix is not provided in the standard output. The focus is on the participation of the documents in topics—comparable with factor scores. If the distribution of a word over the topics is flat, however, this negatively affects the assignment to the top-group of ten words in each topic. Some of the central words in the argument may remain invisible because they are not statistically significant for specific paragraphs.

### *3.3. The Second Case: the "Leiden Rankings"*

In the next round, we upscale to the 687 records retrieved from Google Scholar with the search string ‘"Leiden ranking" OR "Leiden rankings." These docuemnts contain 1,778 words (stop words excluded), which occur 4,724 times. However, this list still contains noise words in languages other than English, such as "und", "der", "de", "la", etc. Sixty-four words occur ten or more times, among which 56 are not stop words. We use these 56 words and repeat the analysis as above.

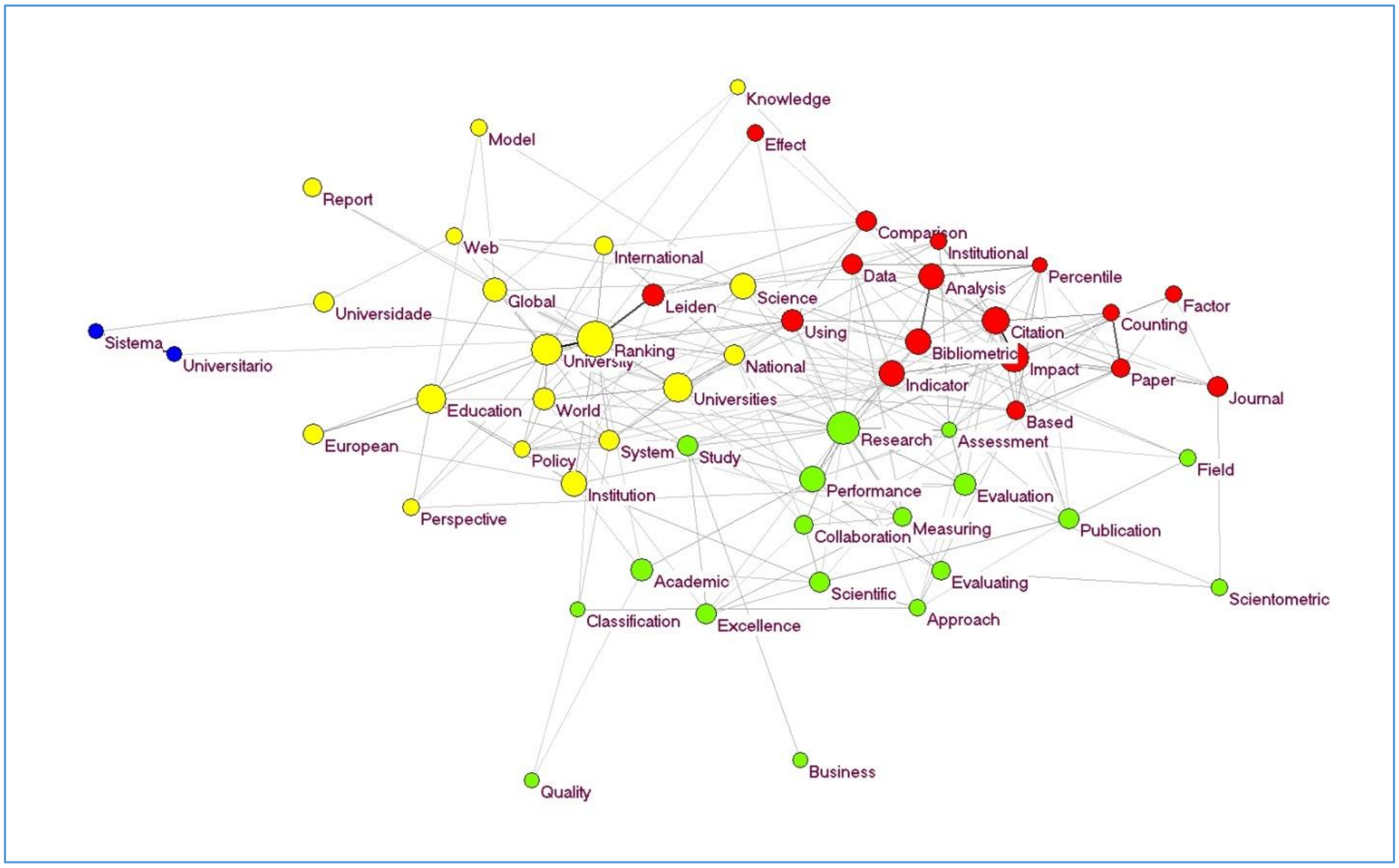

**Figure 6**: Four clusters of 56 words in a cosine-normalized map (cosine > 0.1) distinguished by the algorithm of Blondel *et al*. (2008); modularity $Q = 0.36$. Kamada & Kawai (1989) used for the layout.

Figure 6 shows that the cosine-normalized map of these 56 words is organized even more clearly than the one based on the Leiden Manifesto (Figure 2 above). As can be expected, this network is more loosely organized than a co-word network based on an argument in a single document. A single paper is more densely packed than a set of documents from various authors (Leydesdorff & Hellsten, 2006). Accordingly, the minimal threshold had to be reduced to cosine > 0.1.[14]

[14] With the threshold at cosine > 0.2, 27 words are no longer connected ($Q = 0.751$).

The clusters can easily be recognized as meaningful: two Italian words are set apart (in blue), but in relation to a cluster of words (yellow) that indicate the ranking of universities at the global level. The cluster of red-colored words indicates bibliometric analysis, while the green-colored words are used to discuss the effects of rankings on universities and research. Using factor analysis of the word/document matrix in this case, the scree plot confirms that no more than three or four groups should be distinguished.

Let us now run the same file using topic modeling. We chose the same parameters as before: five topics of ten words. The resulting 50 words indicate several non-English stop words as topical, such as the articles from Latin languages ("la", "las", and "los"). However, 35 words could be matched with the 56 words used above.[15] Figure 7 shows the result of using topics for the coloring of the nodes in Figure 6; Table 3 lists the topics in terms of ten words each. Because some words are attributed to more than a single topic, 39 words can be compared statistically: Cramér's $V$ = .240; $p$ = .811. The two representations are thus significantly different.

**Table 3**: Word lists of five topics in the set of 687 documents, using LDA/MALLET as implemented at https://code.google.com/p/topic-modeling-tool/

| *topicId* | *words..* |
|---|---|
| 1 | rankings higher institutions indicators world science los european excellence und |
| 2 | de la performance en based ranking universidades las espa sciences |
| 3 | universities education ranking academic evaluation indicator systems citations papers policy |
| 4 | impact data study leiden scientific global national comparison international approach |
| 5 | research university citation analysis case bibliometric web fields bibliometrics collaboration |

[15] The word "ranking" is classified in both topic 2 and 3. We assigned it (arbitrarily) to topic 2 in Figure 7.

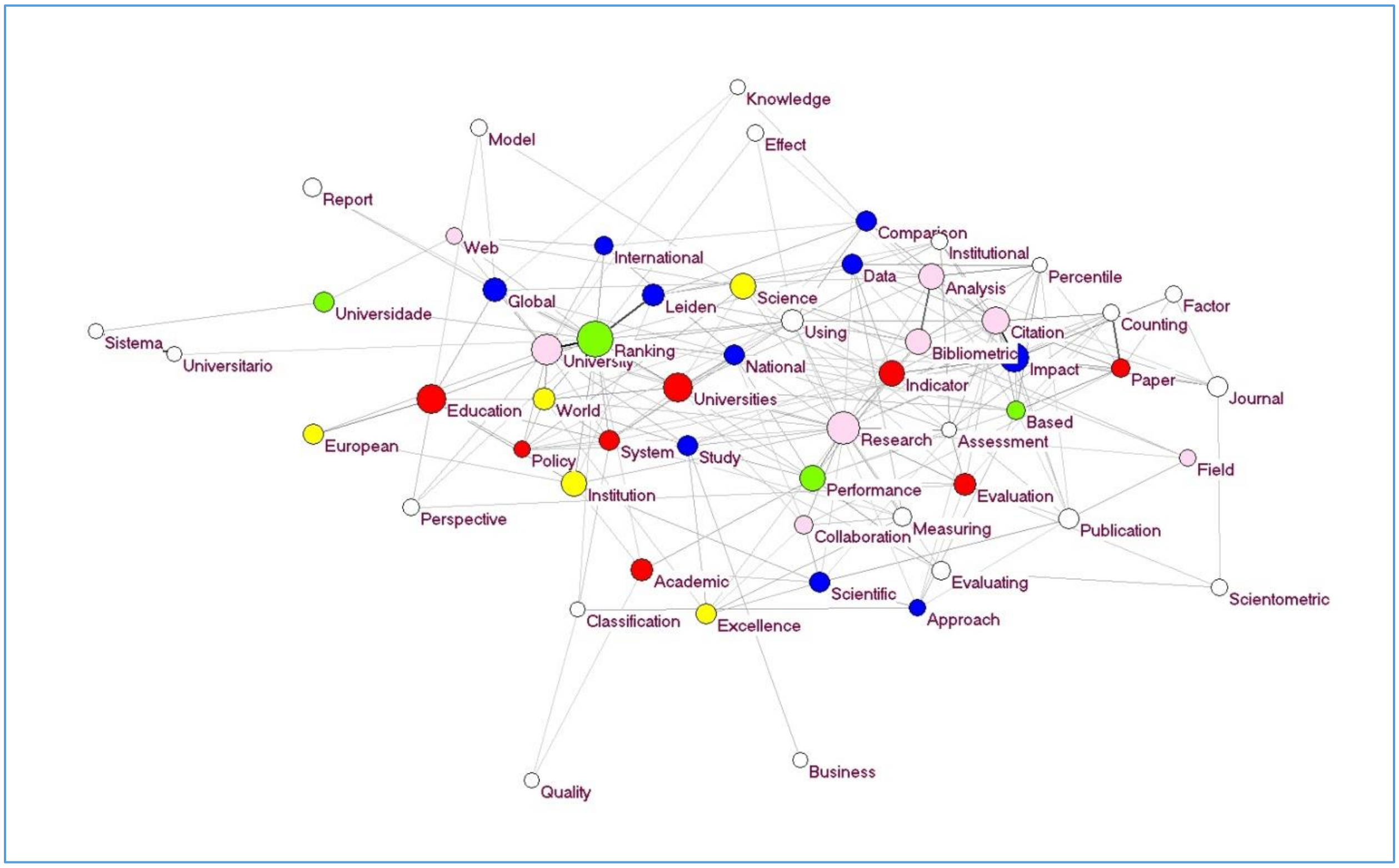

**Figure 7**: Layout as in Figure 6; but coloring of the nodes according to the LDA model. (Words not covered by the LDA output are colored white.)

In our opinion, these results are confusing. Not only are "university", "ranking", and "Leiden" placed in different topics, "research" and "collaboration" (pink-colored) are assigned to topics different from "performance" (green) and "evaluation" (red). Important words such as "publication" (or "publications") and "quality" are not considered topical.

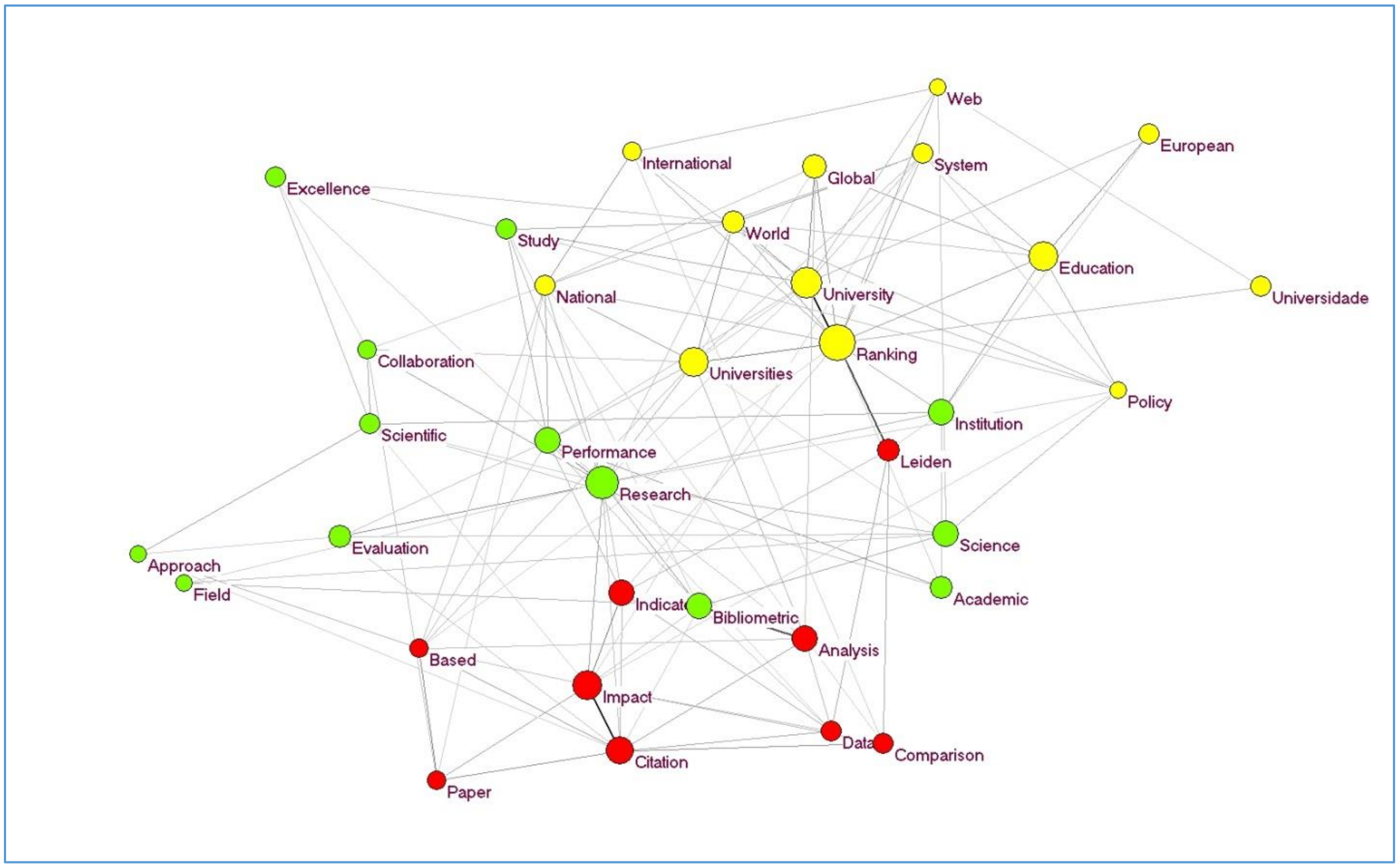

**Figure 8**: Four clusters of 35 "topical" words in a cosine-normalized map (cosine > 0.1) distinguished by the algorithm of Blondel *et al*. (2008); modularity $Q = 0.33$. Kamada & Kawai (1989) used for the layout.

If we extract only the 35 words that were assigned to topics (among these 56 words), and again run the algorithm for community-finding as used above, we are able to reproduce the three main groups distinguished above (Figure 8): (1) global university ranking (in yellow); (2) bibliometric analysis (red); and (3) effects on research (in green). In other words, the topical words can be organized meaningfully, but the topic model did not achieve this. Although we extended the set by more than an order of magnitude, the topic model still distinguished topics on the basis of considerations other than semantics (e.g., statistical or linguistic characteristics). While others have argued that the robustness of LDA topic model results is unaffected by the lack of semantic and syntactic information (Mohr & Bogdanov, 2013), our results suggest differently in the case of small and medium-sized samples (*n* of documents < 1,000).

**Table 4**: Word lists of ten topics in the set of 687 documents, using LDA/MALLET as implemented at https://code.google.com/p/topic-modeling-tool/

| *topicId* | *words..* |
|---|---|
| 1 | research science leiden approach ranking information class medicine scientometrics economy |
| 2 | de la en los espa el fica superior educaci sobre |
| 3 | impact education indicators comparison papers evaluation funding effect evaluating von |
| 4 | institutions performance data universidades las policy web del output para |
| 5 | based academic global systems sciences cient report university der implications |
| 6 | ranking education using indicator rankings highly journal model management knowledge |
| 7 | universities analysis european excellence counting universitarios normalized europe individual cross |
| 8 | rankings world study universities institutional publications bibliometric international und university |
| 9 | university citation top perspective publication quality innovation strategic olas distributions |
| 10 | scientific national fields bibliometrics citation cited collaboration level scholarly countries |

At the suggestion of one of our referees, we also explored a topic model with ten topics, each represented by ten words (Table 4). Fifty-five of the hundred instances can be compared with the five-topics solution: Cramér's $V = .49$; $p = .035$; and thus moderately significant. However, reading the words in the respective topics is not particularly helpful, since one can easily reason a semantic context among these words in the different combinations. Although some readings are counter-intuitive, one can always argue that this alert function is one of the purposes of the analysis. From this perspective, a user evaluation is not sufficient ground for choosing one method over another. However, it remains that the words "performance," "evaluation," and "research" are classified in different topics (that is, topics 4, 3, and 1, respectively).

## Conclusions and discussion

The pervasive problem with grouping words algorithmically is that one can easily rationalize a meaningful coherence among words *ex post*. Validation is therefore a pertinent problem (Grimmer & Stewart, 2013); but sometimes one can compare among results using different approaches (cf. Rafols & Leydesdorff, 2009). In this study, we investigated the claim that "topic modeling" provides an improved alternative to semantic or co-word mapping based on the word/document occurrence matrix. The issue is urgent since programs for topic modeling have become user-friendly and very popular in some disciplines, as well as in policy arenas.

Our argument provides a *caveat* against the use of topic modeling for mapping the semantic structures contained in moderately sized sets of documents ($n$ < 1,000): we were not able to produce a topic model that outperformed the co-word maps, and the results of the two models were not related. The differences between the co-word maps and the topic models were statistically significant, although we used the same input data and the same list of stop words in both cases. Our argument is not at the level of the ensuing visualizations (since these require additional choices such as the layout), but at the level of the underlying statistics. The visualizations serve as illustrations to our argument.

One limitation to our study is the use of small and medium-size document sets and the choice of LDA in the implementation of the topic-modeling software available from Google. In addition to LDA, a number of other algorithms have been developed for topic modeling, such as Hierarchical Dirichlet Process (HDP), etc. However, what these algorithms have in common is that they are based on statistical models developed for machine learning and natural language processing using computers, whereas co-word models have been developed in the information sciences with the objective of a meaningful interpretation of the results (Callon *et al*., 1986; Rip, 1997). As topic models are further developed in order to handle "big data," validation becomes increasingly difficult, since no human interpretation can capture such large volumes of text. However, the computer algorithm may find nuances and differences that are not obviously meaningful to a human interpreter (Chang *et al*., 2010; Jacobi *et al*., 2015, at p. 6).

Additionally, unlike semantic networks, topic models allow no flexibility regarding the selection of words to be fitted to the model. After the correction for stop words, topic models categorize all remaining words into topics. Whereas this particular treatment of data can be useful in reducing human bias, it also limits the extent to which the analyst is able to use contextual knowledge to separate between relevant and irrelevant words for the problem under investigation.

We also compared two co-word models. The fast development of social network analysis over the past twenty years has led to the development of software that can be used in the analysis of co-word maps extracted from texts. Using VOSviewer or Pajek as examples, we have shown that one can use these programs for processing small and medium-sized samples to obtain meaningful and (significantly) comparable results. One (dis)advantage of VOSviewer may be that the program makes some parameter choices in the background. Using the word/document matrix, an analyst is able to countermand these choices. In other words, the choice among these programs can be pragmatic, particularly in the case of large sets. In case of doubt, one can always return to the underlying word/document matrices. While the visualizations can be used to support an argument, one should be careful not to build one's arguments exclusively based on visualizations because maps remain projections on a two-dimensional plane of a multi-dimensional complexity.

In the case of topic modeling, informed choices of parameters (e.g., the number of topics) require testing the model. This is often beyond the competence of a user who wishes to apply a topic model to make or illustrate an argument. The availability of MALLET topic modeling at the Google interface invites, in our opinion, its use without further reflection. As we have shown, the results can then be confusing for a number of reasons. Topic modeling is a statistical technique based on a random starting point, so that one cannot expect two runs to provide the same results. One can run the model a number of times until one finds a satisfying solution, but the validity of that solution is as doubtful as any other solution.

The counter argument here can be that the results of separate runs represent different interpretations of the *same* data, and are therefore compatible by definition (Goldstone & Underwood, 2012). While such an argument might be satisfactory in methodological research settings, this statistical approach does not resolve the validity issues of topic model solutions. Sets of seven or ten words can almost always be rationalized ex post. However, these sets of words may not necessarily represent topics a human reader would identify because topic models uncover statistical patterns within a corpus whether or not they are interpretable (Blei & Lafferty, 2009). The topics remain algorithmic artifacts that do not match with the interpretation of a human reader in the case of small and medium-sized samples. Our results do not exclude the possibility that topic models perform better when larger sets are analyzed. However, the claim that upscaling to large sets would improve the quality of the topics remains unproven.

Users who are unable to read large amounts of texts may be satisfied using topic modeling even if the quality of these indicators is debatable. As noted, the NSF had a model of 1000 topics constructed on the basis of 170,000 awards granted between 2000 and 2012. Using this model for the evaluation, Nichols (2014, at p. 747) claimed that 89% of the awards granted by the directorate of the Social and Behavioral Sciences are indicated as "interdisciplinary research." To what extent is this astonishing result perhaps a consequence of the mixing of disciplinary terminologies by the topic model?

As noted, the qualitative interpretability of topics in terms of words does not inform us about the quality of the clustering of the documents in the set. Topic modeling may be an effective tool for clustering large ("big") sets of documents, comparable with techniques such as community-finding in social network analysis (Blondel *et al*., 2008; Ding, 2011; Newman & Girvan, 2004). However, one cannot infer from the quality of the clustering of documents along the one axis of the word/document matrix to the quality of the grouping of the words along the other axis. In the case of limited document sets, such as the semantic mapping of scholarly discourses in terms of words and co-words, our results give reason for concern.

**Acknowledgement**

We are grateful to Kasper Welbers and the three anonymous referees for comments on an earlier version of this paper.